# Generalization vs. Hallucination

Xuyu Zhang[1,2], Haofan Huang[3], Dawei Zhang[2], Songlin Zhuang[2], Shensheng Han[1,5], Puxiang Lai[3,4, *], and Honglin Liu[1,5, *]

[1]*Shanghai Institute of Optics and Fine Mechanics, Chinese Academy of Sciences, Shanghai 201800, China*
[2]*School of Optical-Electrical and Computer Engineering, University of Shanghai for Science and Technology, Shanghai 200093, China*
[3]*Department of Biomedical Engineering, The Hong Kong Polytechnic University, Hong Kong SAR, China*
[4]*Photonics Research Institute, The Hong Kong Polytechnic University, Hong Kong SAR, China*
[5]*Center of Materials Science and Optoelectronics Engineering, University of Chinese Academy of Science, Beijing 100049, China*
*[puxiang.lai@polyu.edu.hk](mailto:puxiang.lai@polyu.edu.hk), and [hlliu4@hotmail.com](mailto:hlliu4@hotmail.com)*

**Abstract**
With fast developments in computational power and algorithms, deep learning has made breakthroughs and been applied in many fields. However, generalization remains to be a critical challenge, and the limited generalization capability severely constrains its practical applications. Hallucination issue is another unresolved conundrum haunting deep learning and large models. By leveraging a physical model of imaging through scattering media, we studied the lack of generalization to system response functions in deep learning, identified its cause, and proposed a universal solution. The research also elucidates the creation process of a hallucination in image prediction and reveals its cause, and the common relationship between generalization and hallucination is discovered and clarified. Generally speaking, it enhances the interpretability of deep learning from a physics-based perspective, and builds a universal physical framework for deep learning in various fields. It may pave a way for direct interaction between deep learning and the real world, facilitating the transition of deep learning from a demo model to a practical tool in diverse applications.

**Introduction**
Deep learning, as an important branch of artificial intelligence (AI), has achieved remarkable breakthroughs and developments over the past decade, being applied in various fields, such as computer vision, natural language processing, healthcare, financial technology, gaming and entertainment, and environmental science [1-6]. However, generalization remains to be a critical and common challenge in deep learning [7-9]. In the field of optical imaging, the generalization of a neural network is related to various factors, e.g., different datasets, numerous optical systems, and alterations in incident angles, wavelengths and optical elements under the same optical system, posing huge hurdles for the adaptability of the network.

Taking imaging through scattering media [10-12] as an example, a neural network trained under a specific condition typically only performs well under the same circumstance. Once the condition changes, the network often fails to reconstruct the image. The relationship between the input and output of an imaging system can generally be expressed as $Y = TX$, where $X$ is the input target, $Y$ is the output, i.e., a speckle pattern, and $T$ is the scattering transmission matrix of the system. It has been found that the key to address the generalization issue between different datasets is to learn a relationship that closely approximates the mapping of the real physical system, and the solution to achieve this goal is optimizing the training dataset so that each pixel on the input and output planes is sufficiently trained [13]. Other types of generalization challenges can be all attributed to $T$. The ability to identify multiple different matrices $T$s of a neural network, in other words, the number of different inverse mappings $T^{-1}$s a network can accommodate, determines its generalization capability.

Under a specific condition, after training a network learns an approximate mapping relationship $M$, which can approach $T^{-1}$ infinitely in principle [13], to predict the image [14,15]. The network exhibits limited generalization capability near $T$, for example, if $corr(T, T') \geq 0.5$, the network can reconstruct images for both $T$ and $T'$, though for the latter case it typically shows degraded image quality. However, when $T'$ is unrelated to $T$, the network fails to predict images for $T'$ [16]. To enhance the network's generalization across various $T$s, one potential approach is to train the network to simultaneously recognize multiple $T$s, and make them consecutively arranged by proper selections to enhance the generalization. The network′s compatibility to different $T$s determines the upper threshold of its generalization capability. It should be noted that we are not talking about the generalization to predict from unseen scattering media based on residual ballistic light, whose $T$ is constant [16], in this study.

Compared to other systems, lensless imaging through scattering media offers an ideal physical model. In this model, the system's configuration and parameters remain constant, and the transmission matrix $T$ can be altered simply by changing the refractive index distribution within the illuminated region. There is no coupling between different physical variables, while the situation changes for many other cases. For instance, once altering the distance between the camera and the scattering medium, $T$ changes, and so does the field of view, speckle size, image resolution, et al., making it difficult to separate the influences of different factors. In contrast, by varying the refractive index distribution, $T$ changes, while maintaining other factors constant.

By employing a typical CNN network in conjunction with a lensless system of imaging through scattering media, we explored the neural network's ability to accommodate multiple transmission matrices $T$s. Our findings demonstrate that training with multiple $T$s can enhance the network's generalization capability, and uncover the underlying causes of hallucinations, indicating a general solution to improve generalization and overcome hallucinations. This research also enhances the interpretability of deep learning from the preceptive of physics, advances theoretical understanding, and promotes the applications of AI in various fields.

**Methods**
A solid-state laser (MGL-III-532-200mW, Changchun New Industries Optoelectronics Tech) with a wavelength $\lambda = 532nm$ was adopted as the illumination source, whose output was expanded to illuminate a digital micromirror device (DMD, V-7001 VIS, ViALUX). The reflected light from a target displayed on the DMD was scattered by a ground glass diffuser (DG10-120-MD, Thorlabs), and the corresponding speckle pattern was recorded by a CCD camera (Ace acA2440-75um, Basler). The object distance $Z_1 = 16cm$, and the image distance $Z_2 = 10cm$. The DMD has an array size 1024×768 with a pixel size of 13.7 $\mu m$, and the CCD sensor is an array of 2448×2048 with a pixel pitch of 3.45 $\mu m$. Handwritten digit images from the MNIST database [17,18] were used as the targets, which were reshaped into 64×64 arrays and displayed on the center of the DMD in subsequence. The diameter of the collimated beam on the DMD is 5 mm, sufficient to cover the whole region of the targets.

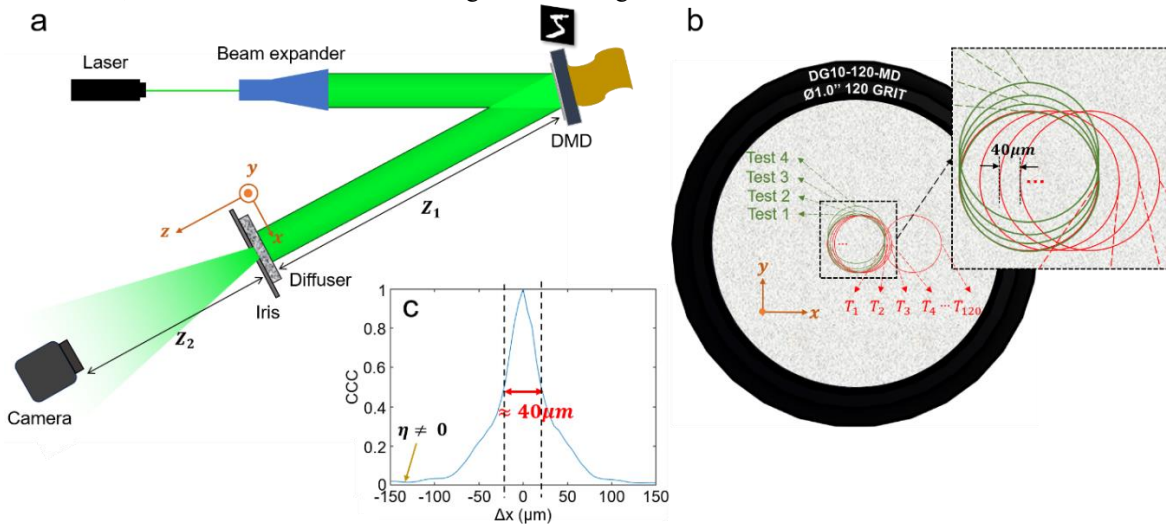

Fig.1 Schematics of the experiment. (a) The experimental setup, the iris has a diameter of 5 mm. (b) Illustration of the arrangement of illumination areas to obtain training and testing data, the circles denote the illumination areas. The dash square is zoomed to see details clearly. $T_{120}$ still has some overlap with $T_1$. (c) The cross-correlation coefficient curve for the diffuser at different displacements along the $x$ direction.

The isoplanatic range of the 120-grit ground glass disk used in experiment is measured to be 40 $\mu m$ (Fig.1c). It means that a 40 $\mu m$ lateral shift of the diffuser results in a completely different transmission matrix, thus, fundamentally altering the mapping relationship of the experimental system. When the illumination region shifts within 20 $\mu m$ from the center, the new mapping relationship is similar to the original one, i.e., $corr(T, T') \geq 0.5$, and the network retains a generalization ability. To test the network's compatibility with different mapping relationships, we collected training data from different illumination regions, varying the number of independent mappings included in each training session to assess the network's compatibility. As shown in Fig.1b, the red circles represent the illumination regions. $T_2$ is shifted 40 $\mu m$ to the right of $T_1$, with each subsequent shift of 40 $\mu m$ until $T_{120}$. This results in a total lateral shift of $(120 - 1) \times 40\mu m = 4.76mm$. For each $T$, the camera captures 5000 speckle images for training (a pair of an input target and the corresponding speckle pattern is a training sample) to ensure the learned relationship stable. The illumination region of Test 1 coincides with $T_1$, while Tests 2-4 are shifted upwards by 10, 20, and 50 $\mu m$ from Test 1, respectively.

As the number of mapping relationships in the training set increases, the data volume, and consequently, the training time, also rises rapidly. Two training strategies were adopted to a trial study.

**Strategy I**: The network started with the same initial conditions, and $5000 * n$ pairs of data, where $n$ was the number of mapping relationships, were used for training at once.

**Strategy II**: Each time a new transmission matrix $T$ was added, and the corresponding 5000 pairs of data were used to continuously train the network. This is to say, the network corresponding to $n$ $T$s was trained by applying the new 5000 pairs of training data on the network already trained for $(n-1)$ $T$s.

Compared to Strategy I, the computational resource consumption in the Strategy II was significantly reduced. In both cases, Test 1 was used for evaluation, and the Pearson Correlation Coefficient (PCC) [19, 20] between the predicted images and the ground truth was used to measure the prediction accuracy.

The generalization range in illumination area of the network was demonstrated and quantified in a paired comparison, where the training and testing areas were modified.

In the study, a CNN (U-Net) was adopted on Tensorflow2/Keras under the hardware environment of NVIDIA RTX3090. The batch size is 10, the time required for each epoch is about 49s, and when the number of training samples doubles, the training time also doubles.

Usually, there is a small amount of ballistic light residue when a light beam passes through a diffuser, and the transmission matrix of the ballistic component does not change with refractive index variations in the illuminated region. In experiment, it is challenging to eliminate ballistic residue or control its proportion. Therefore, in addition to experimental data, we also included simulation data for training and testing, where the proportion of ballistic component could be precisely controlled. Based on wave optics, we simulated the corresponding speckle pattern for each digit image. In simulation, the ground glass was modeled by a random phase screen with a phase function $\exp[-jC\phi(x,y)]$, where the phase $\phi(x,y)$ had a Gaussian random distribution [16, 21], and the proportion $\eta$ of ballistic residue was controlled by adjusting the coefficient $C$.

**Results**

In experiment, a performance comparison between the two training strategies is shown in Fig.2. As the number of $T$ increases, the quality of the predicted images gradually decreases and then stabilizes. The prediction accuracy of the two strategies is comparable, with negligible differences. Therefore, Strategy II was adopted for subsequent network training. Due to the presence of residual ballistic light, the PCC stabilizes at 0.65, enabling successful image prediction even when $n = 40$. Here, the ballistic residue is the key to such generalization capability [16].

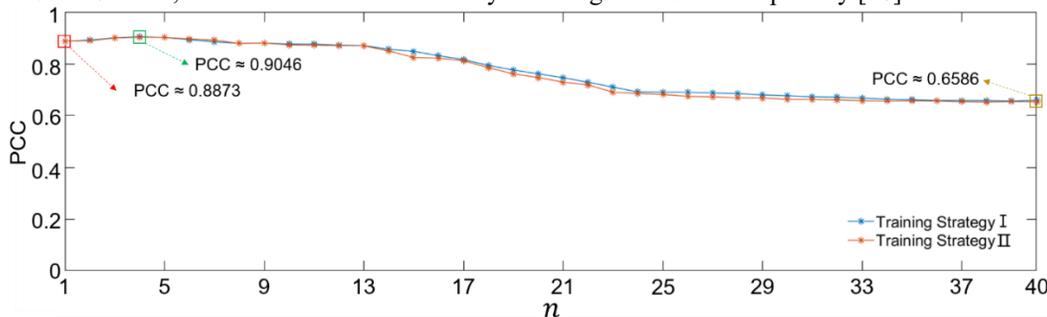

Fig.2 Comparison of Strategies I and II.

Based on the cross-correlation coefficient curve in Fig.1, whose tail is higher than 0, the proportion of residual ballistic light after passing through the diffuser is estimated to be $\eta \approx 0.01$. Fig.3 compares the networks trained with experimental data and simulated data, where two cases of $\eta = 0$ and $\eta = 0.01$ are considered for the latter. When $\eta = 0.01$, the simulation results match the experimental data well, confirming the reliability of the simulation and the proportion estimation of the ballistic residue. Without ballistic residue ($\eta = 0$), as shown by the red squares and the curve, the network's predictive ability diminishes, when the number of $T$ reaches 30, with prediction failures (see the four inserted images for $n = 31, 36, 41, 51$). In contrast, the prediction at $n = 41$ for $\eta = 0.01$ is still recognizable, as shown in the inserted image above the yellow curve, where information is extracted from ballistic light instead of scattered light. From Fig.3, we can say that without ballistic light the network can still predict images well when $n = 20$, that is to say, the network can accommodate 20 independent mappings $T$s. Therefore, the number of $T$s that the U-net network can effectively handle is selected as 20.

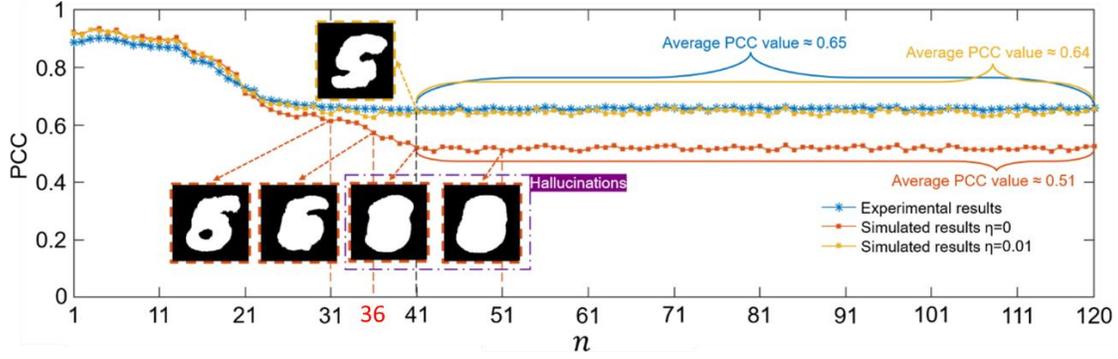

Fig.3 Compatibility results of the U-net network. Experimental results consist well with the simulation results at $\eta = 0.01$.

To assess the network's compatibility with different $T$s for scattered light and demonstrate the enhanced generalization capability, it is necessary to eliminate the influence of ballistic components. Fig.4 shows the predicted imaging results of the network in a paired comparison when it is trained to accommodate 20 different $T$s in the absence of ballistic residue with simulation data. In Group I, $T_{i+1}$ is shifted 40 $\mu m$ to the right of $T_i$, and five different regions were tested. Test A coincides with the region of $T_1$, while Tests B, C, D and E are shifted to the right of $T_1$ by $0.5 \times 40\ \mu m$, $2.3 \times 40\ \mu m$, $9.6 \times 40\ \mu m$ and $18.2 \times 40\ \mu m$, respectively, with the latter four positions randomly selected within the scanning range from $T_1$ to $T_{20}$. The network successfully predicted images through all five positions, demonstrating generalization across the range of $T_1$ to $T_{20}$. In Group II, $T_{i+1}$ was shifted 30 $\mu m$ to the right of $T_i$, and five different regions were tested. Test A coincides with the region of $T_1$, while Tests B, C, D and E were shifted $0.5 \times 30\ \mu m$, $2.3 \times 30\ \mu m$, $9.6 \times 30\ \mu m$, and $18.2 \times 30\ \mu m$ to the right of $T_1$, respectively. Similar to Group I, the network could successfully image through any position within the scan range. However, due to partial correlation between $T_i$ and $T_{i+1}$, the actually accommodated independent mapping relationships was less than 20, which results in better imaging quality compared to Group I. Obviously, by carefully selecting mapping relationships, the network's generalization can be improved, ensuring reliable predictions over a broader range, i.e., enhancing the generalization of the network. To further enhance generalization to various changes in the system, efforts should be focused on modifying the network to increase its compatibility to different mapping relationships.

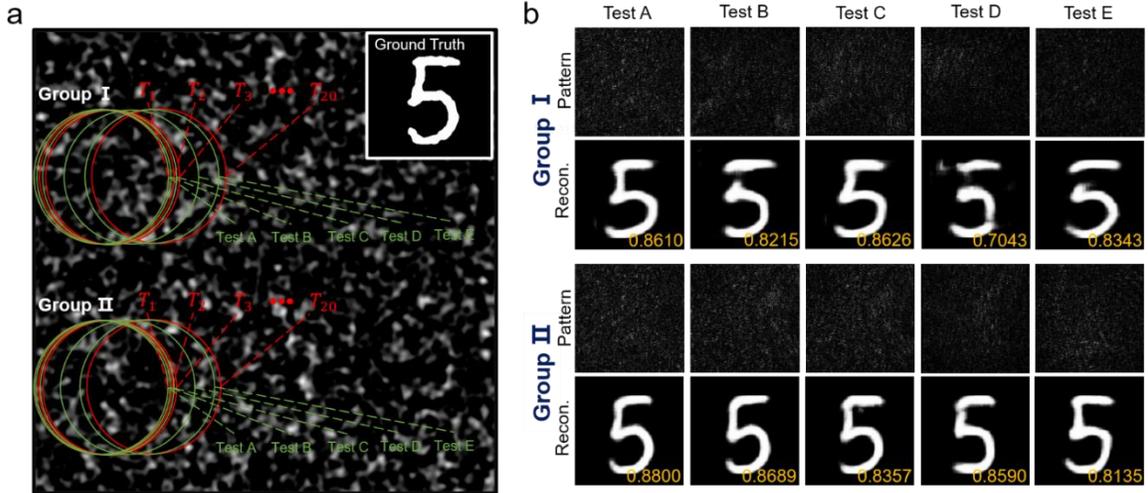

Fig.4 Generalization demonstration of the CNN in imaging through a scattering medium. (a) Schematic illustration of different illumination areas to obtain the training and testing data. The insert in the up right corner is the ground truth. (b) Predicting results of digit target "5" in Tests A-E, the PCC value of each reconstructed image with the ground truth is added in its bottom.

Fig.5 presents the results of network predictions under varying numbers of independent $T$ by shifting the diffuser along $x$ axis (see Fig.1b). Without ballistic residue, the network can still predict images within the permissible number of independent $T$, though the image qualities in Test 1 decline slightly. However, the network's prediction capability rapidly deteriorates from Test 1 to 4. At $\Delta y = 20 \ \mu m$, near the edge of original generalization range, the prediction is merely successful; at $\Delta y = 50 \ \mu m$, the prediction fails (Effective relationships are learned, and there is target information in the speckles, but the network can't decode it due to mismatching between encoding and decoding. If there is no target information in speckles, the same phenomena will appear, too.). When the number of independent $T$ exceeds the network's compatibility, the network must compromise to all $T$s, leading to a failure to approximate any specific $T$. This results in predictions completely unrelated to the ground truths, i.e., hallucinations [22-26], as shown in the fifth and sixth rows of Fig.5a. Clearly, these hallucinations arise due to the network's failure to learn any effective mapping relationship, resulting in the prediction lacking of constraint. Despite this, the network captures a general feature that handwritten digit images are centered, which is why the hallucinations have such characteristic. However, when a small amount of ballistic component is introduced ($\eta = 0.01$), the target becomes recognizable, because the network has learned the mapping relationship of ballistic component and allowed to extract information from the ballistic residue based on that relationship. The hallucination from the scattering component (more specifically due to the compromised relationship learned from scattered light) remains, creating artifacts in the image. Increasing the proportion of ballistic component can reduce and eventually eliminate these artifacts. As an effective mapping relationship emerges, the hallucinations diminish even disappear. The network can simultaneously learn mapping relationships for both the scattering and ballistic components, allowing it to extract information from both. With fewer $T$s, the network primarily extracts information from the scattering component, but as the number of $T$ increases, the network's ability to extract images from the ballistic component is strengthened. Since the ballistic component is almost unaffected by position, the network's sensitivity to positional shifts decreases in such prediction, i.e., with enhanced generalization to see through unseen scattering media [16].

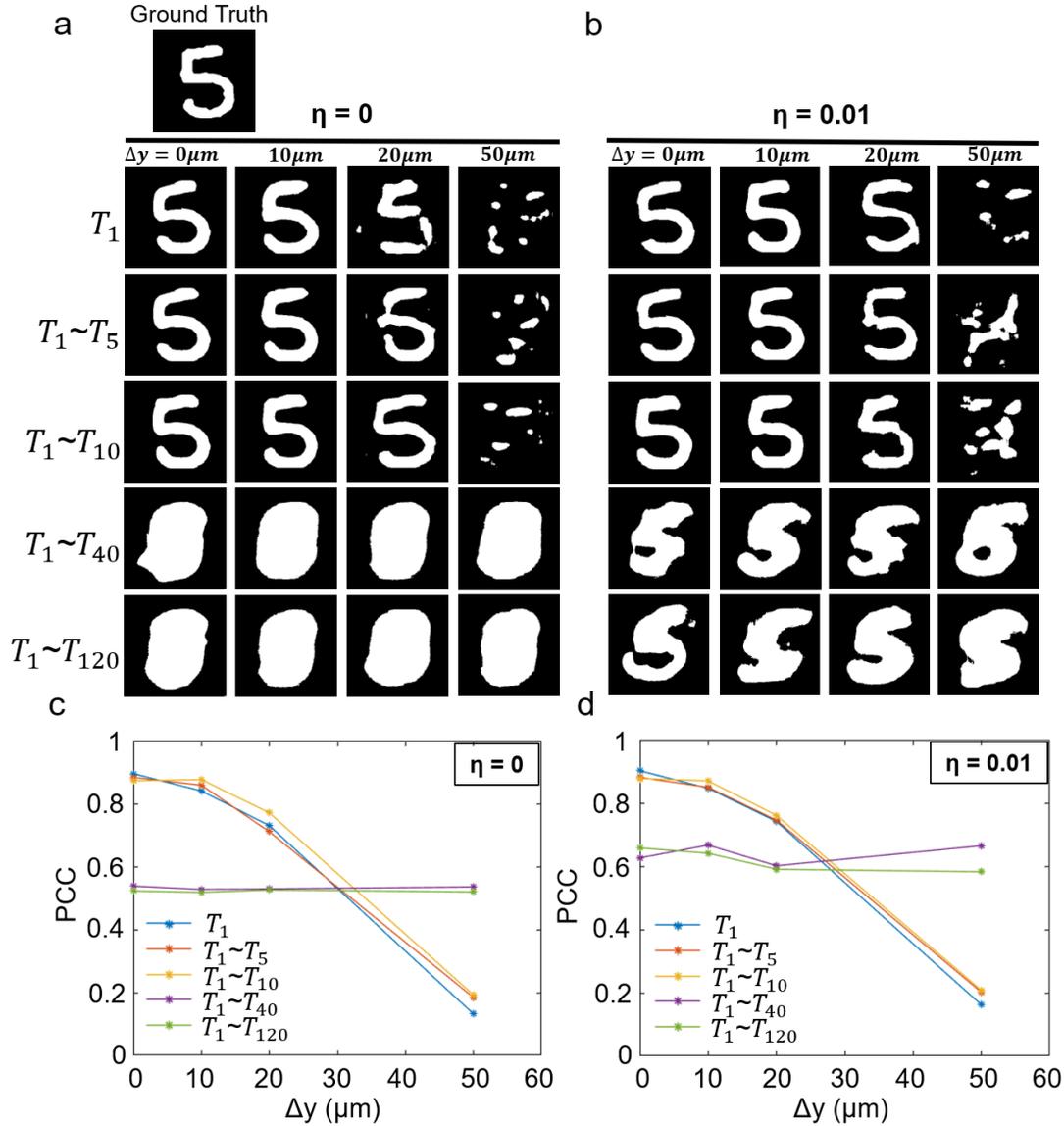

Fig.5 Comparison of image prediction under different numbers of $T$ in training between $\eta = 0$ and $\eta = 0.01$. (a) $\eta = 0$, information is only extracted from scattered light, beyond the compatibility of the network, no image can be reconstructed but hallucinations appear. Within the compatibility, the image can be reconstructed, but the quality deteriorates as $\Delta y$ increases. When $\Delta y = 0$, the image quality for the case of training only with $T_1$ is the best, since there is no compromise between different $T$. (b) $\eta = 0.01$. Within the compatibility, information can be reconstructed from scattered light, as shown in the first three rows of reconstructions at $\Delta y = 0, 10, 20 \ \mu m$. Information can also be extracted from ballistic photons, its influence on reconstruction is trivial due to its small portion, however, can be enhanced as more areas are seen in training. Beyond the compatibility, the network fails to predict the target from scattered photons, but its ability to extract information from ballistic light is enhanced, and images can be recognized even though there are hallucinations generated from scattered light. (c, d) Variations of the PCC value with $\Delta y$ in different cases.

The creation process of a hallucination with increased number of $T$ is illustrated in Fig.6. As the number of $T$ increases in training, the reconstructed images become increasingly distorted until they no longer reserve any feature of the target to recognize it, i.e., the output image is irrelevant to the input target. In a word, the distortion becomes a hallucination when reaching an extreme. The proportion of artifacts in the predicted image for different numbers of $T$ is added below each image. Without ballistic component ($\eta = 0$), neoplasms dominate when the number of $T$ reaches 40 or more, leading to hallucinations. However, introducing a small amount of ballistic component ($\eta = 0.01$) significantly reduces artifacts, allowing the target to be discerned again.

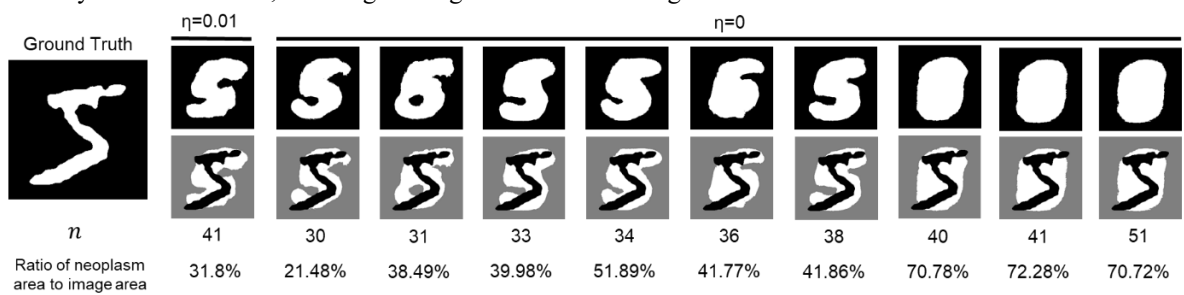

Fig.6 Creation process of a hallucination in the prediction of a digit 5 with increased number of $T$. Reconstructed images are displayed in the first row, the second row shows the overlay results of the digit in black on the reconstructed images. As the ratio of the neoplasm area to the image area goes beyond 70%, a hallucination is created.

**Discussions**
Based on above findings, training a network with data from multiple distinct $T$s allows the network to learn multiple inverse mapping relationships $T^{-1}$s simultaneously. However, as the number of $T$ increases, the learned relationship increasingly deviates from every mapping, eventually turns to be irrelevant to any specific $T^{-1}$. In other words, there is no regulation in the trained network to constrain image recovery, and the prediction is random and aimless, that is, disorder is the cause of hallucinations. Here, generalization and hallucination are inherently associated; improving generalization by accommodating more mapping relationships results in network predictions closer to hallucinations. To enhance generalization while suppressing hallucinations, a straightforward strategy is to expand the network's architecture, providing greater flexibility, which is the solution adopted in big models. However, while enhancing compatibility and improving generalization, it introduces greater complexity in network and unpredictability in performance. The future research of AI, particularly in deep learning and large models, could focus on designing appropriate network architectures to increase generalization capabilities while anticipating and mitigating potential unexpected outcomes.

There are different definitions of the hallucination, after the semantic shift in the late 2010s, the emphasis turns to be the generation of factually incorrect or misleading output by AI systems. About its causes there are speculations on data and modeling, such as, source-reference divergence and imperfect modeling. By studying the generalization in $X$ [13] and $T$, the causes of artifacts/incorrectness in predictions are differentiated. In this paper, hallucinations are incorrect generations due to lack of effective constraints in network prediction, to be more extreme, totally unrelated to the input. Someone might insist the artifacts/incorrectness caused by source-reference divergence as hallucinations, please bear in mind that this type of hallucination can be suppressed by enhanced generalization, i.e., learning a relationship approaching the real mapping as close as possible [13]. Neglecting the immeasurable complexities and infinite source requirements, there is still no way to integrate all principles of nature and rules of humanity into an AI model, since many are unknowns, hence, hallucinations can be mitigated by not eliminated for an open model. It is different for closed spaces, such as chess and Go, where all rules are accurately known, there is no hallucination in network prediction normally.

A neural network is usually considered as a black box, it is difficult to explain how the internal changes affecting the output. In order to enhance interpretability, traditional investigations focus on the conceptual representation of hidden layers and changes in weight parameters, such as hidden layer analysis methods, simulation/agent models and sensitivity analysis methods [27-31]. Different from above approaches, in our study the simulation capability of a CNN network to an imaging system is checked with reference to changes in the real physical system. The generalization capability of a network can be quantified and its performance is predictable. Moreover, it will be possible to locate problems and provide potential solutions. This paves a new way for network interpretability. In future, combing with key parameter change monitoring, sensitivity analysis, statistical analysis, etc., the interpretability of deep learning could be greatly improved, thus build a solid foundation to enhance practical applications of AI in various fields.

## Conclusions
By employing a physical model of imaging through scattering media, the issue of deep learning's lack of generalization to changes in system response functions is investigated. We identified the root causes of this limitation and proposed a universal solution, explained the occurrence of hallucinations and discovered the relationship between hallucination and generalization. In combination with Ref. [13], a general physical framework for applications of deep learning in various fields is built. The interpretability of deep learning is also enhanced from a physical modeling perspective. It may reshape the developments of deep learning, facilitate the transition of deep learning from a demo model to a practical tool in diverse applications, and enable direct interactions between AI and the real world someday.

## Funding
The work was supported by National Natural Science Foundation of China (NSFC) (81930048), Guangdong Science and Technology Commission (2019BT02X105), Hong Kong Research Grant Council (15217721, R5029-19, C7074-21GF), Shenzhen Science and Technology Innovation Commission (JCYJ20220818100202005), and The Hong Kong Polytechnic University (P0045680, P0043485, P0045762, P0049101).

## Acknowledgements
H. L. conceived the idea and designed the study. X. Z. implemented the experiment and simulation. H. L. X. Z., H. H., and P. L. analyzed the data and wrote the manuscript. All contributed to revise the manuscript.

## Competing Interests
The authors declare no conflict of interests.